\documentclass[aps,prd,nopacs,groupedaddress,floatfix,nofootinbib]{revtex4}

\usepackage{graphicx}
\usepackage{bbm}
\usepackage{amsmath}
\usepackage{amssymb}
\usepackage{mathrsfs}
\usepackage{xcolor}
\usepackage{tikz}
\usepackage[normalem]{ulem}
\usepackage[hidelinks]{hyperref}
\DeclareGraphicsExtensions{.eps, .eps, .jpg, .png}

\newcommand{\intd}{\thinspace\mathrm{d}}

\newcommand{\partiald}[2][]{\frac{\partial #1}{\partial #2}}

\newcommand{\reals}{\mathbb{R}}

\newcommand{\lagr}{\mathcal{L}} 

\DeclareMathOperator{\sgn}{sgn}
\DeclareMathOperator{\csch}{csch}
\begin{document}



\title{Instantons for particles joined by strings in three dimensional gravity}


\author{Liam O'Brien$^{1\, 2}$}

\email[]{lobrien@caltech.edu}
\author{Lorenzo Sorbo$^{1}$}

\email[]{sorbo@physics.umass.edu}

\affiliation{$^1$Amherst Center for Fundamental Interactions, Department of Physics, University of Massachusetts, Amherst, MA 01003, U.S.A.\\ $^2$Department of Physics, California Institute of Technology, Pasadena, CA 91125, USA}


\date{\today}

\begin{abstract}

We study the instantons describing the production of particles at the ends of codimension-one objects (strings and struts) in $(2+1)$-dimensional Minkowski and de Sitter spaces.  A Minkowskian background allows only for systems with vanishing total mass, so that either negative mass particles or negative tension struts are required. On a de Sitter background, on the other hand, we find processes describing the production of string/particle systems with no negative energies involved. We also compute the probabilities of creating and of breaking an infinite cosmic string in de Sitter space. We perform our analysis of the system in de Sitter space employing a generalization of the toroidal coordinate system to the three-sphere.

\end{abstract}


\maketitle



\section{Introduction}%

One of the most striking properties of the relativistic quantum theory of fields is the possibility of matter production on nontrivial stationary classical backgrounds. Such nonperturbative processes have been recognized since the early '30s~\cite{Sauter:1931zz} and have been studied with a number of techniques.  Most remarkably, an expanding Universe provides an ideal background for this phenomenon, and the precise measurements of the statistical properties of the Cosmic Microwave Background, along with the predictions from the inflationary scenario, provide good evidence that this kind of process is at the origin of the cosmological structures we inhabit. 

Cosmological perturbations are assumed to originate from the amplification of the vacuum fluctuations of light states. Gravitational production of heavy states has received comparatively less attention, but has nevertheless been the subject of a significant amount of work.  In particular, nonperturbative production of black holes has been the subject of interest since the mid-80's, after the proposal by Gibbons~\cite{Gibbons:1986cq}. Besides the configuration of a pair of black holes at the opposite ends of Euclidean de Sitter space~\cite{Mellor:1989wc,Mann:1995vb}, more complicated constructions involved, for instance, black holes at the end of cosmic strings~\cite{Eardley:1995au,Dias:2003st}, charged black holes at the end of a cosmic string in external electromagnetic fields~\cite{Hawking:1995zn,Emparan:1995je} and black holes in the presence of electromagnetic and a dilaton field~\cite{Dowker:1993bt}.  An extensive review of the literature can be found in~\cite{CamposDias:2003tv}. 

In this paper we focus our analysis on $(2+1)$-dimensional systems. There are a few motivations to do so. To start with, the topological nature of $(2+1)$-dimensional gravity leads to simple, closed form calculations of the Euclidean action for our systems. Moreover, when comparing to the $(3+1)$-dimensional case, point particles and strings/struts in $(2+1)$ dimensions look less like black holes and strings/struts  and more like strings and membranes, which have received comparatively less attention. One additional difference is that, since there is a maximum mass allowed in $(2+1)$-dimensional systems, initial configurations with infinitely long strings (such as those studied in~\cite{Hawking:1995zn,Emparan:1995je}) are not allowed, which limits the space of possible initial conditions. 

In order to set our system up, we start by discussing systems with vanishing cosmological constant. Our starting point is the Wick-rotated three dimensional space in toroidal coordinates~\cite{bateman}, that can be obtained by cutting the $(3+1)$-dimensional Minkowskian C-metric~\cite{Kinnersley:1970zw} with a plane through its axis and by setting the mass parameter to zero. Such a metric describes empty $(2+1)$-dimensional Minkowski space with two foci that follow uniformly accelerating worldlines. As shown in~\cite{Anber:2008zz}, by appropriately cutting and pasting within this geometry one can construct solutions that describe accelerating particles joined by a codimension-$1$ object (that is a string or strut, depending on the sign of its tension, in our $(2+1)$-dimensional spacetime). If we assume our initial space to be empty Minkowski, then either the particles or the string/strut must have a negative mass -- Minkowski space is known to be stable~\cite{Witten:1981mf} as long as no negative-energy objects exist in the theory. 

Our main analysis focuses on de Sitter backgrounds, where matter can be created by the existence of a nontrivial gravitational background, even if it  satisfies positive energy conditions. As a consequence, the de Sitter case allows us to consider a wider set of physically interesting  conditions. These include in particular ``short'', negative tension struts - whose length is less than half of the circumference of Euclidean de Sitter - as well as ``long'', positive tension strings, ending in both cases on positive mass particles. In this study of the de Sitter case we introduce a coordinate system, that one could call ``toroidal on the sphere'', that to our knowledge has never been used in the study of $(2+1)$-dimensional de Sitter space (see~\cite{Pascu:2012yu} for a collection of coordinate systems in de Sitter). We finally compute the probabilities of creating and of breaking a closed cosmic string that circles the equator of $(2+1)$-dimensional de Sitter space.

We use the ``mostly plus'' metric signature, and work in units with $\hbar = c = k_B = 1$, so that the Einstein  equations  read $G_{\mu\nu} = T_{\mu\nu}/M_3$, where $M_3 \equiv 1/8\pi G_3$.  
 

\section{Decay of Minkowski space}
\label{sec:decay_mink}%

In this Section we describe the decay of Minkowski space into two point particles located at the end of a strut, a negative tension codimension-one object.

\subsection{Geometry of the system} \label{geometry_Minkowski}

Let us start by reviewing the geometry describing a system of a strut ending on two positive mass particles in $(2+1)$-dimensional Minkowski space, that was first discussed in~\cite{Anber:2008zz}.  We introduce a dimensionless coordinate system $(t,V,\theta)$, where $t \in \reals$ is timelike, and $V\in\reals$, $\theta\in(-\pi,\pi]$ are  space-like coordinates (with $\theta$ playing the role of an angular coordinate).  In these coordinates, we write three dimensional flat space as
\begin{equation}
\intd s^2 = \dfrac{1}{A^2\left[\cosh V + \cos\theta\right]^2}\left\{-\sinh^2V \intd t^2 + \intd V^2 + \intd \theta^2\right\}\,,
\label{metric_Minkowski}
\end{equation}
where $A$ is a constant with dimensions of a mass (we will describe this coordinate system below, see also~\cite{Anber:2008zz}). Since three-dimensional gravity is not dynamical, we can introduce singular matter sources by cutting-and-pasting this metric. More specifically, we consider the possibility that our flat space decays into a configuration involving a singular stress-energy tensor of the form $T_{\mu\nu} = T\,\delta(\theta)\,h_{\mu\nu}$, where $h_{\mu\nu}$ is the induced metric on the surface $\theta = 0$. This describes a strut of tension $T<0$ located at $\theta = 0$. Note that we do not include the point particles at the end of the strut explicitly in the stress-energy tensor; we shall see later that the particles are necessarily included in the solution in order to conserve energy. 

By imposing (dis)continuity of (the derivatives of) the metric at the strut, we re-write the metric~(\ref{metric_Minkowski}) as
\begin{equation}
\begin{split}
&\intd s^2 = \dfrac{C^2}{A^2\left[\cosh\left(C\,V\right) + \cos\left(C\left(\theta +\sgn(\theta)\delta\right)\right)\right]^2}\left\{-\sinh^2(CV) \intd t^2 + \intd V^2 + \intd \theta^2\right\}\,,\qquad -\pi<\theta\le \pi\,, \\
&\sin(C\delta) \equiv - \frac{T}{2\,A\,M_3}\,,\qquad\qquad C \equiv \frac{\pi}{\pi + \delta}\,,
\end{split}
\label{metric_final}
\end{equation}
where the expression for $C$ is obtained by  requiring the metric to be continuously differentiable as $\theta\to\pm\pi$.

To see the inclusion of the point mass, we note that by taking the asymptotic form of the metric as $V \to \pm\infty$ and defining $r \equiv 2e^{-C\,|V|}/A$, we recover the standard conical deficit metric (up to a scale factor in the timelike coordinate)
\begin{align}
\intd s^2\big|_{|V|\to \infty}\simeq -C^2 \intd t^2/A^2+\intd r^2+C^2\,r^2\,\intd\theta^2\,.
\end{align}
This indicates there are two point masses at $V = \pm\infty$, each with mass $m = 2M_3\,C\delta$. In terms of $m$ and $T$ only, the solution parameters are
\begin{equation}
\begin{split}
&C = 1 - \frac{m}{2\pi M_3}\,,\quad \delta = \frac{\pi m}{2\pi M_3 - m}\,, \\
&A = -\frac{T}{2\,M_3}\,\frac{1}{\sin\left(m/2M_3\right)}\,.
\end{split}
\label{eq:parameters}
\end{equation}
We also define the mass $m_{\text{strut}}$ of the strut as follows: first, we integrate over a $\theta = 0$, $t = \text{constant}$ hypersurface to find the proper length of the strut. Then, we multiply the result by the strut tension $T$. Doing so, we find $m_{\text{strut}} = 2\,T\,C\delta/(A\,\sin(C\delta))$. Using our relations among the model parameters, we find $m_{\text{strut}} = -2m$ -- the total mass of the system is indeed zero. 

To gain some insight into the geometry of the solution~\eqref{metric_final}, consider the map
\begin{equation}
\begin{split}
&\mathcal{T} = \frac{\sinh(CV)}{A[\cosh(CV) + \cos(C(\theta + \sgn(\theta)\delta))]}\sinh(Ct)\,,\\
&\mathcal{X} = \frac{\sinh(CV)}{A[\cosh(CV) + \cos(C(\theta + \sgn(\theta)\delta))]}\cosh(Ct)\,,\\
&\mathcal{Y} = \frac{\sin(C(\theta + \sgn(\theta)\delta))}{A[\cosh(CV) + \cos(C(\theta + \sgn(\theta)\delta))]}\,.
\end{split}
\label{eq:mappings}
\end{equation}

By inverting these transformations and substituting into our metric~\eqref{metric_final}, or substituting this change of coordinates directly into the three-dimensional Minkowski metric $\intd s^2 = - \intd \mathcal{T}^2 + \intd \mathcal{X}^2  +\intd\mathcal{Y}^2$, we find that this coordinate transformation is an isometry between our metric and the Minkowski metric. With appropriate conventions, one can also recognize this coordinate map as the definition of toroidal coordinates in Lorentzian space~\cite{bateman}. In this case, the $\mathcal{X}$ axis is the axis along which the foci (the two point particles) lie. Additionally, the ${\cal T}=t=0$ section describes two dimensional Euclidean space in bipolar coordinates.

\begin{figure}
\includegraphics[scale=.35]{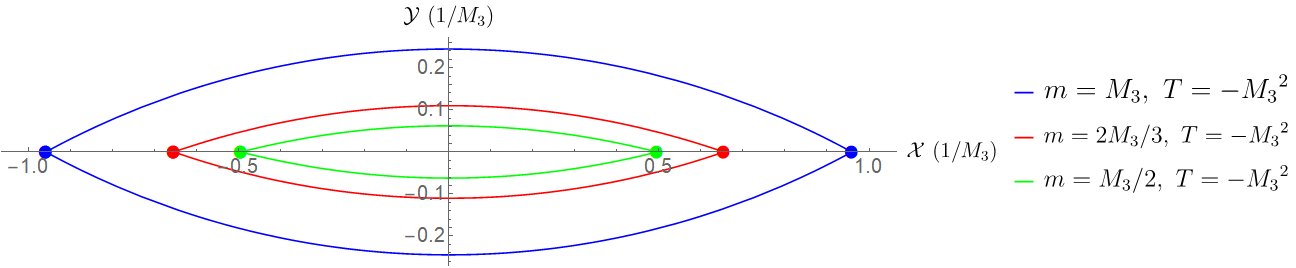}
\caption{The location of the struts and point particles in the locally Minkowski space for $t=0$, as given by the map~(\ref{eq:mappings}), for varying values of the particle mass $m$ and strut tension $T$. The particles and strut for each set of parameters are indicated by the colored points and lines (respectively). Note how the mapping gives two different surfaces for each strut (given in this case by a reflection $y \mapsto -y$ along the strut) - we identify these two surfaces as the same in our space, and effectively cut out the space between them.
The full geometry of the (Euclideanized) space is obtained by rotating the system about the $\mathcal{Y}$ axis, with the (Euclidean) timelike coordinate  $i\mathcal{T}$ orthogonal to the page.}
\label{embedding_Minkowski}
\end{figure}

Using the map~(\ref{eq:mappings}), we can determine the location of the particles and struts. We show the results in Fig.~\ref{embedding_Minkowski} for the constant-time slice $t = 0$ and multiple different particle masses $m$. As indicated in Fig.~\ref{embedding_Minkowski},  each strut ($\theta = 0$ hypersurface) is mapped to two distinct surfaces -- one corresponding to each sign of the parameter $\delta$ in eqs.~\eqref{eq:mappings}. The surfaces $\theta \to 0^+$ and $\theta \to 0^-$ are therefore identified  once the space in between is cut away.

With some further manipulations, we find that the $(\mathcal{T},\mathcal{X},\mathcal{Y})$ coordinates satisfy the conditions~\cite{Anber:2008zz}
\begin{subequations}
\begin{align}
&\mathcal{X}^2 + \left(\mathcal{Y} + \frac{\cot(C(\theta + \sgn(\theta)\delta))}{A}\right)^2 =  \frac{\csc^2(C(\theta + \sgn(\theta)\delta))}{A^2} + \mathcal{T}^2 \label{mapping_circles}\\
&\mathcal{Y}^2 + \left(\sqrt{\mathcal{X}^2 - \mathcal{T}^2} - \frac{\cosh(CV)}{A\sinh(CV)}\right)^2 = \frac{1}{A^2\sinh^2(CV)} \label{mapping_tori}\\
&\mathcal{X}^2 - \mathcal{T}^2 = \frac{1}{A^2}\frac{\sinh^2(CV)}{[\cosh(CV) + \cos(C(\theta + \sgn(\theta)\delta))]^2} \label{mapping_worldlines}
\end{align}
\end{subequations}
Equation \eqref{mapping_circles} shows that constant-$\theta$ surfaces are hyperboloids in the $(\mathcal{T},\mathcal{X},\mathcal{Y})$ coordinates. In particular, if $\mathcal{T}$ is fixed as well, such surfaces form circles parallel to the $\mathcal{X}$-$\mathcal{Y}$ plane. The struts (at time $t=0$) lie on a portion of such a circle, as can be seen in Fig. \ref{embedding_Minkowski}. Additionally, if we Wick-rotate $t \mapsto i\tau$, we see that constant-$\theta$  surfaces form \textit{spheres} in the Euclideanized space.  We also see from \eqref{mapping_tori} that constant-$V$ surfaces form \textit{tori} in the Euclideanized space. Under the Wick rotation, the map~\eqref{eq:mappings} also becomes periodic in  Euclidean time $\tau$ with period $2\pi/C$. We will leverage these facts later to allow for easy computation of the Euclidean action. 

Taking the limit of equation \eqref{mapping_worldlines} as $V \to \pm\infty$ yields the equation of a hyperbola in the $(\mathcal{X},\,\mathcal{T})$ plane: $\mathcal{X}^2 - \mathcal{T}^2  = A^{-2}$. Since $V \to \pm\infty$ corresponds to the positions of the point particles, we conclude the particles are accelerating along the $\mathcal{X}$-axis with proper acceleration $A$ (given in terms of $m$ and $T$ by the last of eqs.~\eqref{eq:parameters}). The surface $V=0$ corresponds to the Rindler horizon for these particles. 

For fixed tension $T$, we see that the acceleration diverges as $m \to 0$, attains a minimum value at $m = \pi M_3$, and diverges again as $m \to 2\pi M_3$ (the maximal magnitude of a point mass in 3 dimensions). In particular, for $m \ll M_3$, the acceleration goes as $A \sim - T/m$.

\subsection{The Euclidean Action} \label{action_Minkowski}

\begin{figure}
\centering
\includegraphics[scale=.4]{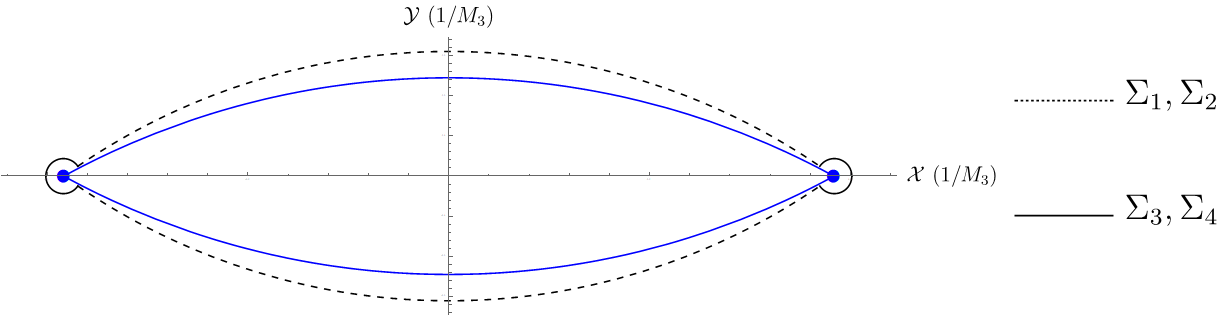}
\caption{The strut and particles (blue) in a $t=0$ slice of our locally Minkowski spacetime, with the boundary surfaces $\Sigma_1$ through $\Sigma_4$ (black) overlaid. $\Sigma_1$ and $\Sigma_2$ are given by the black dashed lines above and below the $\mathcal{X}$-axis, respectively, and $\Sigma_3$ and $\Sigma_4$ are given by the solid black lines above and below the $\mathcal{X}$-axis, respectively. Note that the boundary surfaces for $\mathcal{X} < 0$ and $\mathcal{X} > 0$ are the same, since they differ only by a rotation in (Euclidean) time of $\pi/C$. All surfaces are shown with the parameters $m = M_3$ and $T = -{M_3}^2$, and the boundary surfaces are shown for $\epsilon = 0.1$ and $V_0 = 4.5$.}
\label{boundary_surfaces}
\end{figure}

For our locally Minkowski space, the Ricci scalar is zero away from the sources. Hence, the Euclidean action reads
\begin{equation}
S_{E} = \underbrace{-M_3\left[ \sum_i \int K_{\Sigma_i} \sqrt{h_{\Sigma_i}}\intd ^2 x\right]}_{S_{\text{boundary}}} + \underbrace{\int \lagr_{\text{matter}} \sqrt{g}\intd ^3 x}_{S_{\text{matter}}}\,,
\label{action_grav}
\end{equation}
where the $\Sigma_i$'s are the boundary surfaces of our space~\cite{Gibbons:1977mu} (that we place close to our singular distributions of matter), and $K_{\Sigma_i}$ and $h_{\Sigma_i}$ are respectively the trace of the Euclidean extrinsic curvature and the determinant of the induced metric on $\Sigma_i$ . In order to take care of the singular matter sources in our system, we cut the space along the following surfaces: $\Sigma_1$ and $\Sigma_2$ are the surfaces defined by $\theta =\pm\epsilon$ (respectively), $0<V<V_0$, and $\Sigma_3$ and $\Sigma_4$ are the surfaces defined by $V =V_0$, $\theta\in [\epsilon,\pi]$ and $\theta \in (-\pi,-\epsilon]$ respectively. See Fig.~\ref{boundary_surfaces} for a visualization of these boundary surfaces in our space. We also have to account for the singular lines at the junction of these surfaces -- see Appendix~\ref{Cusp_Action} for a more detailed discussion. The end result for the boundary terms, including the singular lines, and after sending $V_0 \to \infty$, $\epsilon \to 0$, reads~(see Appendices~\ref{app:eh} and~\ref{Cusp_Action} for details)
\begin{equation}
S_{\text{boundary}} = \frac{8\pi M_3\sin(C\delta)}{A\,[1 + \cos(C\delta)]} + \frac{2\pi M_3}{A}\big{(}\pi - 2C\delta\big{)}\,.
\label{boundary_action}
\end{equation}
Likewise, the calculations reported in Appendix~\ref{app:matter} show that the part of action due to the matter yields
\begin{equation}
S_{\text{matter}} = \frac{2\pi\,m}{A} + \frac{2\pi\, T}{A^2\,[1 + \cos(C\delta)]}\,.
\label{matter_action}
\end{equation}
Adding together all the contributions, and using the parameter relations in eqs.~(\ref{eq:parameters}), we finally obtain the action
\begin{equation}
S_E =-\frac{16\pi\, {M_3}^2}{T}\left[ \sin^2\left(\frac{m}{4M_3}\right)+\frac{1}{4} \,\sin\left(\frac{m}{2M_3}\right)\right]\,.
\label{action_final}
\end{equation}
The tunneling probability is given by the exponential of minus\footnote{In general, the probability is the exponential of minus \textit{two} times the action, but the action in eq.~(\ref{action_final}) above is actually twice the action of the instanton, since we have kept in its calculation the entire circle in Euclidean time, whereas the instanton is obtained by joining {\em half} of that circle to the Minkowskian section of the solution.} the difference between the action $S_E$ computed above and the action for empty space (which we denote by $S_0$). Here, $S_0$ is given by the limit of \eqref{action_final} as $m,T \to 0$ in such a way that $A\approx -T/m$ is finite,  so that the metric \eqref{metric_final} remains regular. We find that $S_0 = 2\pi^2M_3/A$, so that
\begin{align}
S_E - S_0=-\frac{16\pi\, {M_3}^2}{T}\sin^2\left(\frac{m}{4M_3}\right)\,,\qquad 0\le m<2\pi\,M_3\,.
\label{action_final_subtr}
\end{align}
For $m \ll M_3$, the difference of the action and background reduces to:
\begin{equation}
S_E - S_0 = - \frac{\pi {m}^2}{T} + \mathcal{O}({m}^4)
\label{action_small_m}
\end{equation}
This yields, to leading order, a tunneling probability of $P \propto \exp(-\pi {m}^2/|T|)$ (since $T<0$). This result is analogous to the one obtained for $(1+1)$-dimensional Schwinger effect, $P\propto \exp(-\pi {m}^2/|eE|)$, if we remind that $-eE$ is, in the regime $|E|\gg e$, the change in energy density in the electric field induced by the pair nucleation~\cite{Schwinger}.


\section{De Sitter background}%

In this Section we generalize the calculation on flat space to the case of de Sitter background. As we will see, in this case the total mass need not be conserved, so that configurations where both the string tension and the particle masses are positive  are allowed. Let us start by describing a coordinate system that is more appropriate to study this system.

\subsection{Toroidal coordinates for three dimensional de Sitter space} \label{bipolar}

To facilitate our discussion of the geometry of our configuration in locally de Sitter space, we construct here a generalization of toroidal coordinates to three-dimensional de Sitter space. Cutting a three-dimensional toroidal system with a plane through the foci gives a plane covered by bipolar coordinates.  Therefore, we will start with a two-dimensional sphere that can be obtained as a section of three-dimensional de Sitter space, and we will look for a generalization of bipolar coordinates to that two-sphere.

Bipolar coordinates for the plane can be constructed starting from the equipotential surfaces of an electric dipole. We use the same procedure to generate bipolar coordinates on the sphere. Consider a two-sphere of radius $\ell$,  equipped with standard angular coordinates $(\vartheta,\,\varphi)$ and, to start with, uniform electric charge density $\sigma$. By symmetry, one can choose the associated electrostatic potential to take the form $\Phi = \Phi(\vartheta)$. Then, a particular solution to the Poisson equation is $\Phi = -Q\log(1 - \cos\vartheta)/4\pi$, with $Q = 4\pi \ell^2 \sigma$ the total charge distributed on the sphere. Note that we choose the solution to be singular at $\vartheta= 0$: this way, there is a point charge of charge $-Q$ at the north pole of the sphere, as dictated by the requirement that this compact space be globally  neutral. Now we rotate the coordinate system $(\vartheta,\,\varphi)$ by an angle $\alpha$ about the $x$-axis to a new one $(\vartheta',\varphi')$.  Under such a rotation $\cos\vartheta \mapsto -\sin\alpha\,\sin\vartheta'\,\sin\varphi' + \cos\alpha\,\cos\vartheta'$, and the point charge now has (Cartesian) coordinates $(0,\,\ell\,\sin\alpha,\,\ell\,\cos\alpha)$. Finally, we add a distribution of charge density $-\sigma$ with a  point charge of opposite charge $+Q$ located at $(0,\,-\ell\,\sin\alpha,\,\ell\,\cos\alpha)$. This way we obtain, after dropping the primes from $\vartheta$ and $\varphi$, the potential on the two-sphere generated by a dipole and with vanishing uniform charge density
\begin{equation}
\Phi = -\frac{Q}{4\pi}\log\left(\frac{1 + \sin\alpha\,\sin\vartheta\,\sin\varphi - \cos\alpha\,\cos\vartheta}{1 - \sin\alpha\,\sin\vartheta\,\sin\varphi + \cos\alpha\,\cos\vartheta}\right)
\label{V}
\end{equation} 
We will use $\Phi$ as one of the coordinates on the two-sphere. To construct the second coordinate, we consider circles on the sphere that cross the foci located at $(0,\,\pm \ell\,\sin\alpha,\,\ell\,\cos\alpha)$. Such circles are obtained by intersecting the sphere with a plane across the foci, the equation for which is found to be $\sin\vartheta\,\cos\varphi + c(\cos\vartheta - \cos\alpha) = 0$. The parameter $c$ that determines the plane (and thereby the circle) and will serve as our second coordinate.

It is convenient to choose the parameters $\theta \equiv \arccos\{1/\sqrt{c^2\,\sin^2\alpha + 1}\}$ and $V \equiv -2\pi \Phi/Q$ to be new coordinates on the sphere.  We then have:
\begin{equation}
\theta = \arccos\left\{\frac{\cos\alpha - \cos\vartheta}{\sqrt{\sin^2\vartheta\,\cos^2\varphi\,\sin^2\alpha+(\cos\alpha - \cos\vartheta)^2}}\right\}\,,\qquad
V = \frac{1}{2}\log\left(\frac{1 + \sin\alpha\,\sin\vartheta\,\sin\varphi - \cos\alpha\,\cos\vartheta}{1 - \sin\alpha\,\sin\vartheta\,\sin\varphi + \cos\alpha\,\cos\vartheta}\right)
\label{mapping_spherical}
\end{equation} 
that can be inverted to:
\begin{equation}
\vartheta= \arccos\left(\frac{\cos\theta + \cos\alpha\,\cosh V}{\cos\theta\,\cos\alpha + \cosh V}\right)\,,\qquad \varphi = \arctan\left(\frac{\sinh V}{\sin\theta}\right)
\label{mapping_spherical_inverse}
\end{equation} 

Substituting these definitions into the standard metric on $S^2$, we find:
\begin{equation}
\intd s^2 = \frac{1}{A^2[(\cos\alpha)^{-1}\cosh V + \cos\theta]^2}\left\{ \intd V^2 + \intd \theta^2\right\}
\end{equation}
where we have defined $A \equiv (\ell\tan\alpha)^{-1}$.

We can now readily extend this analysis to the 3-sphere by introducing a third hyperspherical coordinate $\psi$ such that $\intd s^2 = \ell^2\big{(} \intd \vartheta^2 + \sin^2\vartheta\, \intd \varphi^2 + \sin^2\vartheta\,\sin^2\varphi\, \intd \psi^2\big{)}$. Using eq.~\eqref{mapping_spherical_inverse}, the 3-sphere metric in terms of $V$ and $\theta$ is found right away as
\begin{equation}
\intd s^2 = \frac{ \sinh^2V \intd \psi^2 + \intd V^2 + \intd \theta^2}{A^2\left[\dfrac{\cosh V}{\cos\alpha} + \cos \theta\right]^2}\,.
\label{bipolar_metric_3d}
\end{equation}

The coordinates $(V,\theta,\psi)$ are an extension of the toroidal coordinates to the three-sphere, and form a chart for three-dimensional Euclidean de Sitter space, with metric given by~\eqref{bipolar_metric_3d}. The cosmological constant $\Lambda>0$ is related to the parameters $\alpha$ and $A$ via
\begin{align}
\cos\alpha\equiv \frac{1}{\sqrt{1+\Lambda/(A^2M_3)}}\,.
\end{align}

Note that for $\Lambda=0$ (i.e., for $\alpha=0$), and after analytically continuing $\psi=it$, we recover the metric in toroidal coordinates of eq.~(\ref{metric_Minkowski}).

\subsection{Adding a string/strut: geometry of the system}

We can now add to our de Sitter geometry a string or a strut ending on two particles.

\subsubsection{Adding a strut}

The metric~(\ref{bipolar_metric_3d}) is very close in form to  the Minkowskian coordinate system in eq.~(\ref{metric_Minkowski}). Therefore we can introduce a strut ending on two particles by performing the same steps as in Section~\ref{sec:decay_mink}. Namely, we replace
\begin{equation}
V\to C\,V\,,\qquad \theta\to C\,\theta\,,\qquad \psi\to iCt\,
\label{replacements_dS}
\end{equation}
in the metric~\eqref{bipolar_metric_3d}, and we cut-and-paste at the location of the singular codimension-one object. Doing so, we find the metric:]
\begin{equation}
\intd s^2 = \frac{C^2}{A{}^2[\beta \cosh(CV) + \cos(C(\theta + \sgn(\theta)\delta))]^2}\bigg{\{}-\sinh^2(CV) \intd t^2 + \intd V^2 + \intd \theta^2\bigg{\}}\,,\qquad \beta \equiv \sqrt{1 + \frac{\Lambda}{A{}^2M_3}}\,.
\label{metric_dS}
\end{equation}
As before, we also have $C = \pi/(\pi + \delta)$ and $\sin(C\delta) = -T/2M_3A$. Consequently, we find the same deficit angle at $V \to \pm\infty$ as in the Minkowski case; relations~(\ref{eq:parameters}) hold in the de Sitter case as well. While most of the parameters retain their form from the Minkowski case, the mass of the strut (defined as the strut tension times its proper length) \textit{does not}, and is given by
\begin{equation}
m_{\text{strut}} = \frac{4T}{A\sqrt{\sin^2(C\delta) + \dfrac{\Lambda}{A{}^2M_3}}}\arctan\left(\frac{\sqrt{1 + \Lambda/A{}^2M_3}-\cos(C\delta)}{\sqrt{\sin^2(C\delta) + \dfrac{\Lambda}{A{}^2M_3}}}\right)\,.
\label{strut_mass_dS}
\end{equation}
We see that a non-zero value of the cosmological constant changes the proper length of the strut relative to its Minkowski counterpart. As a result, it is not true in general that the total mass of the system is zero. For the case $\Lambda = 0$, the mass above reduces to the Minkowski result $m_{\text{strut}} = 2TC\delta/A\sin(C\delta)$. In the limit $T\to 0$ with finite $\Lambda$, on the other hand, the mass~(\ref{strut_mass_dS}) converges to $\pi\,T\sqrt{M_3/\Lambda}$ (since $A\to 0$ as $T\to 0$). In particular, this means the Euclidean length of the strut in this limit is half the circumference of de Sitter space (whose radius is $\ell = \sqrt{M_3/\Lambda}$), and that the point particles lie on opposite poles of the Euclideanized space.

\begin{figure}
\includegraphics[scale=.45]{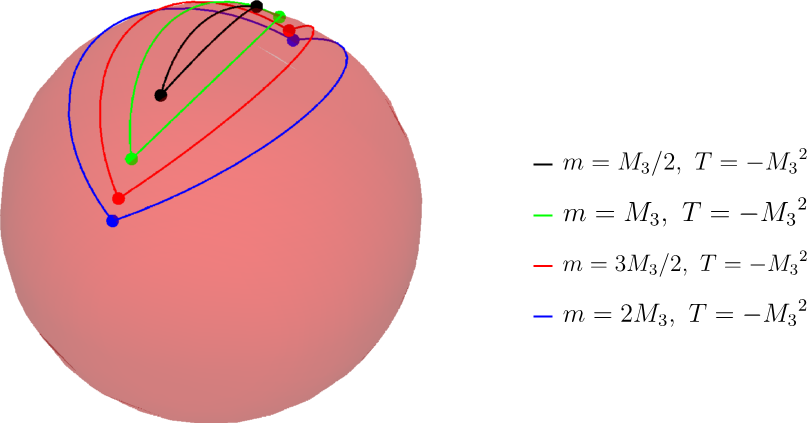}
\caption{A $t=0$ surface (red sphere) in our locally de Sitter space with the embedded particles and struts (points and lines, respectively), as given by~\eqref{mapping_spherical_inverse}, for varying values of the particle mass $m$ and strut tension $T$, and $\Lambda = {M_3}^3$. Note how the mapping~\eqref{mapping_spherical_inverse} yields a different position for the strut for each sign of $\delta$ (corresponding to a shift $\varphi \mapsto \varphi + \pi$) - we identify these two surfaces as the same in our space, and effectively cut out the space between the solid lines in the figure.}
\label{embedding_dS}
\end{figure}

Inspection of the isometry between the metric \eqref{metric_dS} and the standard 3-sphere metric given via the coordinate maps \eqref{mapping_spherical_inverse} with replacements \eqref{replacements_dS} shows that our configuration is periodic in time with period $2\pi/C$. We also can use this coordinate mapping to determine the location of the struts and particles in the underlying (Euclideanized) spacetime. We do so in Fig.~\ref{embedding_dS} for a $t = 0$ slice, for varying values of $m$ and $T$, and fixed $\Lambda$. As with the Minkowski case, the strut ($\theta  = 0$) is mapped to two distinct hypersurfaces - one for each sign of the shift $\delta$. These surfaces are identified after cutting away the space in between. Consequently, the induced hyperspherical coordinates $(\vartheta,\varphi,\psi)$ \textit{do not} cover the entire 3-sphere in the presence of matter. This can be visualized from Fig.~\ref{embedding_dS} by cutting out the patch of the sphere between two lines of the same color, and identifying these two lines with each other and with the location of the strut.

\subsubsection{Adding a string}

Unlike the locally flat case discussed in the previous Section, a de Sitter background allows for a configuration of a positive tension string ending on positive mass particles. Such a situation can be obtained if the string is longer than half of the circumference of the de Sitter sphere, a configuration that is realized by placing the string along the $\theta=\pm\pi$ surface in the coordinate system~(\ref{bipolar_metric_3d}). The metric takes the form
\begin{equation}
\begin{split}
&\intd s^2=\frac{C^2}{A^2\big[\sqrt{1+\Lambda/(M_3A^2)}\cosh CV+\cos(C\theta)\big]^2}\big\{-\sinh^2(CV)\,\intd t^2+\intd V^2+\intd \theta^2\big\}\,,\qquad -\pi<\theta<\pi\,,\\
&\sin(C\pi)=\frac{T}{2A\,M_3}\,,\qquad C=1-\frac{m}{2\pi M_3}\,.
\end{split}
\label{eq:string_metric}
\end{equation}
The latter two relations show that, in this case, $T$  and $m$ have the same sign.
%

\subsection{The Euclidean action}

For locally de Sitter space, there is a non-zero Ricci scalar $R = 6\Lambda/M_3$ away from the sources. Hence, in addition to the boundary surface action and matter action, we have a non-zero Einstein-Hilbert contribution to the action, $S_{EH}$, for our space, that we find to be
\begin{equation}
S_{EH} = -4\pi\frac{M_3^{3/2}}{\Lambda^{1/2}} \left\{\pi-2 \tan^{-1}\left(\frac{(\beta -1) \tan \left(\frac{1}{2} C \delta\right)}{\sqrt{\beta ^2-1}}\right)+\frac{\sqrt{\Lambda}}{A\,\sqrt{M_3}}\frac{\sin (C\delta)}{\beta  (\beta +\cos (C\delta))}\right\}\,.
\label{EH_action_dS}
\end{equation}
Following an analogous procedure as in the Minkowski case, the boundary term action and matter action are (see the Appendices for details)
\begin{equation}
S_{\text{bound}} = \frac{8\pi M_3\sin(C\delta)}{A\,\beta[\beta+\cos (C\delta)]} + \frac{2\pi M_3}{A\beta}\left(\pi - 2C\delta\right)\ , \hspace{10pt} S_{\text{matter}} =  \frac{2\pi\, m}{A\beta} + \frac{2\pi\, T}{A^2\beta[\beta +  \cos C\delta]}   \,.
\label{boundary_matter_action_dS}
\end{equation}

Using the relations~\eqref{eq:parameters} and introducing the dimensionless variables
\begin{align}\label{eq:defxy}
x \equiv \frac{m}{2M_3}\,,\qquad y \equiv \frac{T}{2\sqrt{\Lambda M_3}}\,,
\end{align}
the total Euclidean action eventually takes the form
\begin{equation}
 S_E = -4\pi^2\frac{M_3^{3/2}}{\sqrt{\Lambda}}+8\pi\frac{M_3^{3/2}}{\sqrt{\Lambda}}\,\arctan\left(\frac{\sqrt{y^2+\sin^2x}+y}{\cos x+1}\right)\,,
\label{action_final_dS}
\end{equation}
where the empty  de Sitter result $S_0^{\rm dS} =-4\pi^2\frac{M_3^{3/2}}{\sqrt{\Lambda}}$ is obtained for $x,\,y\to 0$. Note that eq.~(\ref{eq:defxy}) implies that $0 \leq x < \pi$. For negative tension struts, we have also $y<0$ so that in the limit $x\to 0$ (that is, for small particle masses and finite strut tension) we obtain $S_E= S_0^{\rm dS} +\pi {m}^2/|T| + \mathcal{O}({m}^4)$. This is in agreement with the Minkowskian result~\eqref{action_small_m}, as expected since in this case the strut is  much shorter than the de Sitter radius.  

\subsection{The no-strut limit}

Unlike the case of a system on a Minkowski background, it is possible to eliminate the strut from the system while keeping finite mass pointlike particles if the cosmological constant does not vanish. To do so we note that eqs.~(\ref{eq:parameters}) allows for a vanishing tension strut if $A\to 0$. While the Minkowskian metric~(\ref{metric_final}) is singular in this limit, the de Sitter metric~(\ref{metric_dS}) is not, and goes to
\begin{eqnarray}
\intd s^2\Big|_{A\to 0} = \frac{C^2\,M_3}{\Lambda}\frac{1}{\cosh^2(CV)}\bigg{\{}-\sinh^2(CV) \intd t^2 + \intd V^2 + \intd \theta^2\bigg{\}}\, ,
\label{metric_dS_A_to_0}
\end{eqnarray}
which can be brought to a standard de Sitter form with the definitions
\begin{equation}
\begin{split}
&r\equiv\frac{\ell}{\cosh CV}\,,\qquad \tau\equiv C\,\ell\,t\,,\qquad \tilde\theta=C\,\theta\,,\qquad \ell\equiv \sqrt{\frac{M_3}{\Lambda}}\\
&\intd s^2\Big|_{A\to 0}=-\left(1-r^2/\ell^2\right)\,d\tau^2+\frac{dr^2}{1-r^2/\ell^2}+r^2\,d\tilde\theta^2\,,\qquad \qquad 0\leq \tilde\theta<{2\pi}\,{C}\,.
\end{split}
\end{equation}
This equation shows that the system now contains just two particles of mass $m=2\pi M_3(1-C)$.  One is located at $r=0$, and the other at the opposite pole of de Sitter space (which is not covered by this coordinate system).

The action is obtained readily by sending $y \propto T \to 0$ in eq.~(\ref{action_final_dS}) and gives: 
\begin{equation}
 S_E=S_0^{\rm dS}+ 2\pi\,m\,\frac{M_3^{1/2}}{\sqrt{\Lambda}}\,.
\label{action_dS_nostrut}
\end{equation}
This result is consistent with the assumption that the instanton action is equivalent to (minus) the entropy of the state, which in turn is $1/4$ of the de Sitter horizon area  (circumference) in units of $G_3=(8\pi\,M_3)^{-1}$. In fact, given the deficit angle generated by the mass $m$, the de Sitter horizon area is ${\cal A}=2\pi\,\ell\,\left(1-\frac{m}{2\pi M_3}\right)$, so that the action is
\begin{align}
S_E\Big{\rvert}_{T\to 0}=-\frac{{\cal A}}{4\,G_3}=-\frac{8\pi\,M_3}{4}\,2\pi\ell\left(1-\frac{m}{2\pi M_3}\right)=-4\pi^2\,M_3\ell+2\pi m\ell\,.
\end{align}

This thermodynamical interpretation is also consistent with a de Sitter temperature $T_{dS}=1/(2\pi\ell)$, leading to a Boltzmann factor $e^{-m/T_{dS}}$ for the probability of creation of a particle of mass $m$ in a de Sitter patch.

\subsection{Breaking of a string in de Sitter space}

One additional interpretation of our results is that the metric~(\ref{eq:string_metric}) describes the final state after the breaking of a closed string circling the entire de Sitter space. In this case the initial state, given by a string of tension $T>0$, can be constructed as follows.

Consider the metric on the sphere
\begin{align}
\intd s^2=\ell^2\left(\intd\vartheta^2+\sin^2\vartheta\,\intd\phi^2+\sin^2\vartheta\,\sin^2\phi\,\intd\psi^2\right)\,,
\end{align}
and cut it at some angle $\vartheta=\vartheta_0$ to put a positive tension string there. To express the value of $\vartheta_0$ as a function of the tension $T$ one can impose Israel's junction conditions at the surface $\vartheta=\vartheta_0$. We further require this surface to be a surface of symmetry of the metric, so that the metric is symmetric under the replacement $\vartheta\to 2\,\vartheta_0-\vartheta$. This leads to the relation
\begin{align}
\tan\vartheta_0=\frac{2\,M_3}{T\,\ell}\,.
\end{align}
Using this result, we can compute the action of the system, obtaining
\begin{align}
S_E^{\rm string}=-8\pi\,\frac{M_3^{3/2}}{\sqrt\Lambda}\,\arctan\left(\frac{1}{y}\right)\,,
\end{align}
where, as above, $y=\frac{T}{2\sqrt{\Lambda M_3}}$, and that is indeed equivalent to the action~(\ref{action_final_dS}) in the limit $x\to 0$.

The probability of breaking a string in de Sitter space is thus given by the exponential of the difference:
\begin{align}\label{eq:action_breaking}
S_E -S_E^{\rm string}=-8\pi\frac{M_3^{3/2}}{\sqrt{\Lambda}}\,\left[{\rm arctan}\left(\frac{\sqrt{y^2+\sin^2x}-y}{1-\cos x}\right)-{\rm arctan}\left(\frac{1}{y}\right)\right]
\end{align}
which, in the limit $\Lambda\to 0$ and $m\ll M_3$, converges to $S_E -S_E^{\rm string}=\pi\,m^2/T$, in agreement with the result obtained, in the same regime, in the four-dimensional case~\cite{Eardley:1995au}.

As a byproduct, we also find that the probability of nucleating a string in a de Sitter background can be obtained as the exponential of the negative of
\begin{align}\label{eq:action_dsstring}
S_E^{\rm string}-S_0^{\rm dS}=8\pi\,\frac{M_3^{3/2}}{\sqrt\Lambda}\,\arctan\left(\frac{T}{2\sqrt{\Lambda\,M_3}}\right)
\end{align}
that for low tension strings $T\ll \sqrt{\Lambda\,M_3}$ gives the probability $\sim e^{-4\pi\ell^2 T}$.

\section{Discussion and conclusions}%

We have presented exact solutions to the Einstein equations describing pairs of particles connected by a string or strut, for both locally Minkowski (previously found in~\cite{Anber:2008zz}) and de Sitter space. We have also shown explicitly how they are isometric to the standard Minkowski metric and standard 3-sphere metric for Euclidean de Sitter space, with patches of the spaces removed and their boundaries identified at the location of the codimension-one objects. Our solutions are useful for describing accelerated pairs of particles in $2+1$ dimensions. Additionally, to the best of our knowledge, the isometry \eqref{mapping_spherical} between our solution and the 3-sphere is also a new coordinate chart for $(2+1)$-dimensional de Sitter space.

We have also computed the Euclidean action for our solutions, which is associated to the entropy of the corresponding state, such that the exponential of such action can be interpreted as the probability of realizing that state~\cite{Gibbons:1977mu}. In this sense, our main results are given in eq.~(\ref{action_final_subtr}) for Minkowskian backgrounds and in eq.~(\ref{action_final_dS}) for a de Sitter background. Our results have allowed us to determine the actions for the creation of a string extending for a super-horizon length in de Sitter space~(\ref{eq:action_dsstring}) and for the breaking of such a string~(\ref{eq:action_breaking}).

It is worth asking what properties analogous solutions in locally Anti de Sitter (AdS) space would have. While in this work we focused on backgrounds of vanishing and positive cosmological constant, solutions in locally AdS space offer an additional rich set of possibilities. Such solutions may also provide new insights in the context of AdS/CFT correspondence, making them well worth future exploration.   

\acknowledgments We thank David Kastor for several useful discussions, especially on the bipolar/toroidal coordinates in Euclidean space. This work was partially supported by the US-NSF grants PHY-1520292 and PHY-1820675.

\appendix

\section{Einstein-Hilbert action and boundary terms}%
\label{app:eh}

\subsection{Locally Minkowski $(\Lambda = 0)$}

For locally Minkowski space, the Ricci scalar is identically zero away from the matter, so only the boundary terms contribute to the geometric part of the action. We cut the space on the four surfaces $\Sigma_1$ through $\Sigma_4$ described in Section \ref{action_Minkowski}.

As discussed in Section \ref{geometry_Minkowski}, the constant-$\theta$ hypersurface $\Sigma_1$ is a (portion of a) sphere in our Euclideanized space; it obeys the equation $x^2 + y^2 + z^2 = R^2$. The unit normal vector to such a surface is then $n^{\mu} = -(x,\, y,\, z)/\sqrt{x^2 + y^2 + z^2}$ (note we take the \textit{inward}-pointing normal, since this is the direction pointing out of our Euclideanized spacetime). The trace of the extrinsic curvature, which is the divergence of this normal, is then readily found to be $-2/R$. In our case, $R = [A\sin(C(\epsilon + \delta))]^{-1}$ by \eqref{mapping_circles}, so we find that $K_{\Sigma_1} = -2A\sin(C(\epsilon + \delta))$. The induced (Euclidean) metric on the surface $\Sigma_1$ is  $\intd s^2 = C^2\big{\{}\sinh^2(CV) \intd \tau^2 + \intd V^2\big{\}}/A^2[\cosh(CV) + \cos(C(\epsilon + \delta))]^2$; the resulting determinant is  $h_{\Sigma_1} = C^4\sinh^2(CV)/A^4[\cosh(CV) + \cos(C(\epsilon+\delta))]^4$. The contribution to the Euclidean action from $\Sigma_1$ is then
\begin{align}
S_{\Sigma_1} &= 2A \sin(C(\epsilon+\delta))\, M_3 \int_0^{2\pi/C}\int_0^{V_0}\frac{C^2|\sinh(CV)|}{A^2[\cosh(CV) + \cos(C(\epsilon+\delta))]^2}\intd V \intd \tau \notag\\
&= \frac{4\pi M_3 \sin(C(\epsilon+\delta))}{A}\left[\frac{1}{1 + \cos(C(\epsilon+\delta))} - \frac{1}{\cosh(CV_0) + \cos(C(\epsilon+\delta))}\right]\,. \label{Sigma_1_action}
\end{align}
The contribution of $\Sigma_2$ to the action is identical, since the entire solution is even in $\theta$.

In our Euclideanized ($\mathcal{T}\mapsto -i \mathscr{T}$) spacetime, we know from \eqref{mapping_tori} that $\Sigma_3$ forms a portion of torus that is radially symmetric about the $\mathcal{Y}$-axis. To leverage this, let us define a cylindrical coordinate system ($\mathcal{R},\gamma,\mathcal{Y}$) via $\mathcal{X} = \mathcal{R}\cos\gamma$, $\mathscr{T} = \mathcal{R}\sin\gamma$, and introduce the shorthand $\mathcal{R}_0 \equiv \coth(CV_0)/A$ and $\mathscr{R} \equiv \csch(CV_0)/A$. Then \eqref{mapping_tori} reads $\mathcal{Y}^2 + (\mathcal{R}- \mathcal{R}_0)^2 = \mathscr{R}^2$. The unit normal vector in these coordinates is
\begin{equation}
n^{\mu} = (n^{\mathcal{R}},\, n^{\gamma},\, n^{\mathcal{Y}}) = \frac{(\mathcal{R}-\mathcal{R}_0,\, 0,\, \mathcal{Y})}{\sqrt{\mathcal{Y}^2 + (\mathcal{R} - \mathcal{R}_0)^2}}\,.
\label{Sigma_3_normal}
\end{equation}
The normal is correctly oriented, since increasing $\mathcal{R}$ corresponds to increasing $|V|$, which is oriented towards the particles (and out of our space). Taking the divergence of this normal, we find the trace of the extrinsic curvature to be
\begin{align}
K_{\Sigma_3} &= \frac{2\mathcal{R}-\mathcal{R}_0}{\mathcal{R}\mathscr{R}}= 2A\sinh(CV_0) - A\coth(CV_0)[\cosh(CV_0) + \cos(C(\theta+\delta))]\,.
\label{Sigma_3_K}
\end{align}
The induced (Euclidean) metric on the surface $\Sigma_3$ is $\intd s^2 = \big{\{}\sinh^2(CV_0) \intd \tau^2 + \intd \theta^2\big{\}}/A^2[\cosh(CV_0) + \cos(C(\theta + \delta))]^2$, which has determinant  $h_{\Sigma_3} = \sinh^2(CV_0)/A^4[\cosh(CV_0) + \cos(C(\theta+\delta))]^4$. The contribution to the action from $\Sigma_3$ is thus
\begin{align}
S_{\Sigma_3} &= -\frac{M_3}{A}\int_0^{2\pi/C}\left(2\sinh^2(CV_0)\int_{\epsilon}^{\pi}\frac{\intd \theta}{[\cosh(CV_0) + \cos(C(\theta + \delta))]^2} - \coth(CV_0)\int_{\epsilon}^{\pi}\frac{\intd \theta}{\cosh(CV_0) + \cos(C(\theta+\delta))}\right)\intd \tau\notag\\
&= \frac{8\pi M_3\coth\left(\frac{CV_0}{2}\right)}{A}\left(\frac{\pi}{2}-\arctan\left[\tan\left(\frac{C(\epsilon+ \delta)}{2}\right)\tanh\left(\frac{CV_0}{2}\right)\right]\right) + \frac{4\pi M_3}{A}\frac{\sin(C(\epsilon+\delta))}{\cosh(CV_0) + \cos(C(\epsilon+\delta))} \,.\label{Sigma_3_action}
\end{align}

The contribution of $\Sigma_4$ to the action is identical, since the entire solution is even in $\theta$. The total gravitational action is then twice the sum of \eqref{Sigma_1_action} and \eqref{Sigma_3_action}, and the action of the two cusps (see Appendix~\ref{Cusp_Action}).

\subsection{Locally de Sitter ($\Lambda > 0$)}

For locally de Sitter space, we have a constant scalar curvature $R = 6\Lambda/M_3$ everywhere; we thus have a non-zero Euclidean Einstein-Hilbert action $S_{\text{EH}} = -\frac{M_3}{2}\int R+\int\Lambda$ given by
\begin{align}
S_{\text{EH}} =& -2\Lambda\int_0^{2\pi/C}\int_{-\pi}^{\pi}\int_{-\infty}^{\infty} \frac{|\sinh(CV)|}{A^3[\beta\cosh(CV) + \cos(C(\theta + \sgn(\theta)\delta))]^3}\intd V\intd \theta \intd \tau  \notag\\
=&-4\pi\frac{{M_3}^{3/2}}{\sqrt{\Lambda}}\bigg{\{}\pi - 2\arctan\left(\frac{(\beta-1)\tan\left(\frac{C\delta}{2}\right)}{\sqrt{\beta^2-1}}\right) +\sqrt{\beta^2-1}\frac{\sin(C\delta)}{\beta[\beta + \cos(C\delta)]}\bigg{\}}\,. \label{action_EH_dS}
\end{align}

We then need the boundary terms, for which we follow a similar procedure to the Minkowski case. For the surface $\Sigma_1$, the (unit) normal vector is (remaining in the ($t,V,\theta$) coordinates this time) $n^{\mu} = -A[\beta\cosh(CV) + \cos(C(\theta + \delta))](0,0,1)$; the associated extrinsic curvature is then $K_{\Sigma_1} = -2A\sin(C(\epsilon+\delta))$ as in the Minkowski case. The contribution to the action from $\Sigma_1$ is then
\begin{align}
S_{\Sigma_1} &= 2A \sin(C(\epsilon+\delta))\, M_3 \int_0^{2\pi/C}\int_0^{V_0}\frac{C^2|\sinh(CV)|}{A^2[\beta \cosh(CV) + \cos(C(\epsilon+\delta))]^2}\intd V \intd \tau \notag\\
&= \frac{4\pi M_3 \sin(C(\epsilon+\delta))}{A}\left[\frac{1}{\beta[\beta + \cos(C(\epsilon+\delta))]} - \frac{1}{\beta[\beta\cosh(CV_0) + \cos(C(\epsilon+\delta))]}\right]\,. \label{Sigma_1_action_dS}
\end{align}
Also in this case, $\Sigma_2$ gives a contribution to the action identical to that of $\Sigma_1$.

As for $\Sigma_3$, we find the unit normal vector to be $n^{\mu} = A[\beta\cosh(CV)  +\cos(C(\theta + \delta))](0,1,0)$. The associated trace of the extrinsic curvature is $K_{\Sigma_3} = 2A\beta\sinh(CV_0) - A\coth(CV_0)[\beta\cosh(CV_0) + \cos(C(\theta+\delta))]$, and so the contribution to the action is
\begin{align}
S_{\Sigma_3} &= -\frac{M_3}{A}\int_0^{2\pi/C}\left(2\beta\sinh^2(CV_0)\int_{\epsilon}^{\pi}\frac{\intd \theta}{[\beta\cosh(CV_0) + \cos(C(\theta + \delta))]^2} - \coth(CV_0)\int_{\epsilon}^{\pi}\frac{\intd \theta}{\beta\cosh(CV_0) + \cos(C(\theta+\delta))}\right)\intd \tau\notag\\
\begin{split}
&=\frac{2\pi M_3\cosh(CV_0)(2-3\beta^2 + \beta^2\cosh(2CV_0))}{A(\beta^2\cosh^2(CV_0) - 1)^{3/2}}\left[\frac{\pi}{2}-\arctan\left(\frac{(\alpha\cosh(CV_0) - 1)\tan\left(\frac{C(\epsilon+\delta)}{2}\right)}{\sqrt{\alpha^2\cosh^2(CV_0) - 1}}\right)\right]\\
&\qquad + \frac{4\pi M_3 \beta\sinh^2(CV_0)}{A(\beta^2\cosh^2(CV_0)-1)}\frac{\sin(C(\epsilon+\delta))}{\alpha\cosh(CV_0) + \cos(C(\epsilon+\delta))}\,.
\label{Sigma_3_action_dS}
\end{split}
\end{align}
Once again, $\Sigma_4$ gives an equal contribution to the action. The total geometric action is then the sum of eq.~\eqref{action_EH_dS} plus twice the sum of eq.~\eqref{Sigma_1_action_dS} and eq.~\eqref{Sigma_3_action_dS}, plus the action of the two cusps (that is computed in Appendix~\ref{Cusp_Action} below). 

\section{Action from Matter}
\label{app:matter}

Consider the following matter Lagrangian for a point particle of mass $m$ at $V = V_0$ and $\theta=0$ and a strut of  tension $T$ at $\theta = 0$
\begin{equation}
\lagr_{\text{matter}} = m\, \frac{\delta(V-V_0)\,\delta(\theta)}{\sqrt{g_{VV}g_{\theta\theta}}} + T\frac{\delta(\theta)}{\sqrt{g_{\theta\theta}}}\,.
\label{Lagrangian_general}
\end{equation}
Note that the Lagrangian for our spacetime is recovered in the limit $V_0\to\infty$. 

\subsection{Locally Minkowski ($\Lambda=0$)}

In the case of locally Minkowski space, the action from matter reads
\begin{align}
S_{\text{matter}} &= \int_0^{2\pi/C}\int_{-\pi}^{\pi}\int_0^{\infty} \left( m\,\frac{\delta(V-V_0) \delta(\theta
)\,C|\sinh(CV)|}{A[\cosh(CV) + \cos(C(\theta + \sgn(\theta)\delta))]} + T\frac{\delta(\theta)C^2|\sinh(CV)|}{A^2[\cosh(CV) + \cos(C(\theta + \sgn(\theta)\delta))]^2}\right)\intd V \intd\theta\intd \tau \notag \\
&= \frac{2 m}{AC}\left[\pi - 2\arctan\left(\tan\left(\frac{C\delta}{2}\right)\tanh\left(\frac{CV_0}{2}\right)\right)\right] + \frac{2\pi T}{A^2[1+\cos(C\delta)]} \,.
\label{matter_action_Minkowski_temp}
\end{align}
Sending $V_0 \to \infty$, and using \eqref{metric_final}, we find
\begin{equation}
S_{\text{matter}} = \frac{2\pi \,m}{A} + \frac{2\pi T}{A^2[1 + \cos(C\delta)]}\,.
\label{matter_action_Minkowski}
\end{equation}

\subsection{Locally de Sitter ($\Lambda>0$)}

In the case of locally de Sitter space, the action from matter is very similar:
\begin{align}
S_{\text{matter}} &= \int_0^{2\pi/C}\int_{-\pi}^{\pi}\int_0^{\infty} \left( m\frac{\delta(V-V_0) \delta(\theta)\,C\,\sinh(CV)}{A[\beta \cosh(CV) + \cos(C(\theta + \sgn(\theta)\delta))]} + T\frac{\delta(\theta)C^2\,\sinh(CV)}{A^2[\beta\cosh(CV) + \cos(C(\theta + \sgn(\theta)\delta))]^2}\right)\intd V \intd\theta\intd \tau \notag \\
&= \frac{2 m}{AC}\frac{\sinh(CV_0)}{\sqrt{\beta^2\cosh^2(CV_0)-1}}\left[\pi - 2\arctan\left(\frac{(\beta\cosh(CV_0)-1)\tan\left(C\delta/2\right)}{\sqrt{\beta^2\cosh^2(CV_0)-1}}\right)\right] + \frac{2\pi T}{A^2\beta[\beta+\cos(C\delta)]}\,.
\label{matter_action_dS_temp}
\end{align}
Sending $V_0 \to \infty$, and using \eqref{metric_dS}, we find
\begin{equation}
S_{\text{matter}} = \frac{2\pi\, m}{\beta A} + \frac{2\pi T}{A^2\beta[\beta + \cos(C\delta)]}\,.
\label{matter_action_dS}
\end{equation}

\section{Action from Cusps}%
\label{Cusp_Action}

We denote as ``cusps'' the one-dimensional surfaces at the junction of the surfaces $\Sigma_1$ and $\Sigma_3$, and at the junction of the surfaces $\Sigma_2$ and $\Sigma_4$. These cusps can be seen in Fig.~\ref{boundary_surfaces} as the points where the dashed and solid lines for the boundary surfaces meet just off the $\cal{X}$-axis. These cusps can be obtained as the zero-radius limit of a two dimensional curved surface, which we describe below.

\begin{figure}
\centering
\begin{tikzpicture}
\draw[very thick] (0,0) -- (0,5);
\draw (0,5) node[above] {$\mathcal{Y}$};
\draw (5,0) node[right] {$\mathcal{X}$};
\draw[very thick] (0,0) -- (5,0);

\draw[dashed] (0,0) -- (1.9,1.9);
\draw (1.9,1.9) -- (4,4);
\draw[thin] (.75,0) arc[radius = .75,start angle = 0, end angle = 45];
\draw (.75,.287) node[above] {\small $\alpha$};

\draw[red] (2,2) arc[radius = 1.12, start angle = 127.5, end angle = 270];
\draw[thin] (2.704,0) -- (2.704,.56) node[right] {$R$} -- (2.704,1.12);
\draw[thin] (2,2) -- (2.352,1.56) node [above right] {$R$} -- (2.704,1.12);
\filldraw (2.704,1.12) circle (1pt) node[right] {$C$};
\filldraw (2,2) circle (1pt) node[above] {$p_1$};
\filldraw (2.704,0) circle (1pt) node[below] {$p_2$};

\draw[dashed] (0,0) --  (2.704,1.12);
\draw[thin] (1.2,0) arc[radius = 1.2,start angle = 0, end angle = 22.5];
\draw (1.109,.459) node[below right] {\small $\alpha/2$};

\end{tikzpicture}
\caption{The geometry of the cusp. The two intersecting lines (which are approximating the boundary surfaces $\Sigma_i$ in the vicinity of the cusp) are separated by an angle $\alpha$ and joined by a circular section (red) of radius $R$ and center $C$ which is tangent to the lines at points $p_1$ and $p_2$, respectively. The cusp is then obtained by taking the limit of this geometry as $R \to 0$.}
\label{cusp}
\end{figure}
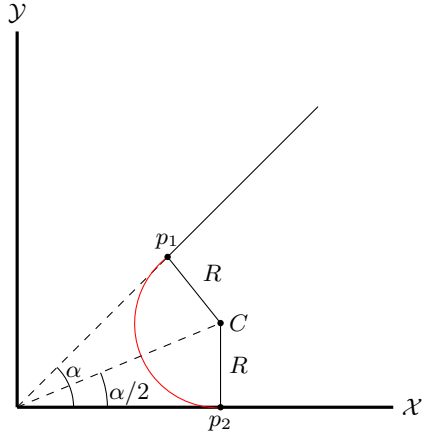

First, we use the cylindrical symmetry about the ${\cal Y}$ axis of the system to focus only on what happens on the $({\cal X},\,{\cal Y})$ plane. Second, we approximate the curves $\Sigma_1,\,\dots,\,\Sigma_4$ on the $({\cal X},\,{\cal Y})$ plane and near the cusp as straight lines. Then, we regularize the cusp as a circular section of radius $R$ that is tangent to the two curves whose intersection generates the cusp. We will take the limit $R\to 0$ at the end of our calculation to obtain the cusp. Let us denote by $p_1$ and $p_2$ the points where the curves intersect the circle. Let us also take the circle to be tangent to the curves at $p_1$ and $p_2$, so that the normal vector is continuous at these points. We further denote by $\alpha$ the angle between the lines, and we initially take one of the lines to be the along ${\cal X}$ axis. This geometry is shown in Fig.~\ref{cusp}.

The coordinates of the circle's center, in this coordinate system, are $C = \left( R\,{\rm {cotan}}\left(\alpha/2\right),R\right)$. To get to a general configuration we perform a rotation of angle $\theta$ and a translation: ${\cal X} \mapsto {\cal X}\,\cos\theta + {\cal Y}\,\sin\theta + {\cal X}_0$ and ${\cal Y} \mapsto -{\cal X}\,\sin\theta + {\cal Y}\,\cos\theta + {\cal Y}_0$. In these more general coordinates, the center of the circle has coordinates ${\cal X}_C = {\cal X}_0 + R\cos\theta\thinspace\cot(\alpha/2) - R\sin\theta$ and ${\cal Y}_C = {\cal Y}_0 + R\sin\theta\thinspace \cot(\alpha/2) + R\cos\theta$. The circle itself is defined by $({\cal X}-{\cal X}_C)^2 + ({\cal Y}-{\cal Y}_C)^2 = R^2$.

We now introduce a cylindrical coordinate system $({\cal X},\,\varphi,\,{\cal Y})$ - with metric $\intd s^2 = \intd {\cal X}^2 + \intd {\cal Y}^2 + {\cal X}^2 \intd \varphi^2$ - where the $\varphi$ coordinate has period $2\pi$, and corresponds to $C$ times the Euclidean time $\tau$. On the circle, ${\cal Y}$ is a function of ${\cal X}$ given by ${\cal Y} = {\cal Y}_C \pm \sqrt{R^2 - ({\cal X}-{\cal X}_C)^2}$; the induced metric $h_{\mu\nu}$ is therefore
\begin{equation}
\intd s^2 = \left(1 + \left(\partiald[{\cal Y}]{{\cal X}}\right)^2\right) \intd {\cal X}^2 + {\cal X}^2\intd \varphi^2\,. 
\end{equation}
The determinant of the induced metric evaluates to $\det h = R^2\,{\cal X}^2/(R^2 - ({\cal X}-{\cal X}_C)^2)$. Finally, in this coordinate system, the unit vector $n^{\mu} = (n^{\cal X},\, n^{\phi},\, n^{\cal Y})$ normal to the circle has components $({\cal X}-{\cal X}_C,\, 0,\, {\cal Y}-{\cal Y}_C)/\sqrt{({\cal X}-{\cal X}_C)^2 + ({\cal Y}-{\cal Y}_C)^2}$. Taking the divergence, we find the extrinsic curvature for this surface to be $K = (2{\cal X} - {\cal X}_C)/R{\cal X}$.

Putting everything together, the action for this hypersurface is
\begin{align}
S_{\text{cusp}} &= -2\pi M_3\int_{{\cal X}_1}^{{\cal X}_2}\frac{2{\cal X} - {\cal X}_C}{\sqrt{R^2 - ({\cal X}-{\cal X}_C)^2}}\sgn({\cal X})\intd {\cal X} \notag\\
&= -2\pi M_3\sgn({\cal X}) \bigg{\{}{\cal X}_C\arctan\left(\frac{{\cal X}-{\cal X}_C}{|{\cal Y}-{\cal Y}_C|}\right) - 2|{\cal Y}-{\cal Y}_C|\bigg{\}}\bigg{\rvert}_{{\cal X}_1}^{{\cal X}_2} \,,\label{action_temp}
\end{align}
where we have already performed the trivial integral in $\varphi$. The limits of integration can be determined by finding the coordinates of $p_1$ and $p_2$ and applying the rotation and translation; we obtain ${\cal X}_1 = R\cot\alpha(1+\cos\alpha) + {\cal X}_0$ and ${\cal X}_2 = R\cot(\alpha/2)\cos\theta + {\cal X}_0$. Using these limits to evaluate \eqref{action_temp}, and taking the limit $R \to 0$, we finally obtain:
\begin{equation}
S_{\text{cusp}} = -2\pi M_3|{\cal X}_0|\times\left[\begin{cases} \pi - \alpha -\theta & 0 \leq \alpha+ \theta \leq \pi/2\\  \alpha + \theta & \pi/2 < \alpha+ \theta \leq \pi\\ 2\pi - \alpha- \theta & \pi < \alpha+ \theta < 3\pi/2\\ \alpha+ \theta - \pi & 3\pi/2 < \alpha+ \theta < 2\pi\end{cases}\ - \begin{cases} \theta & 0 \leq \theta \leq \pi/2\\  \pi -  \theta & \pi/2 < \theta \leq \pi\\ 2\pi - \theta & \pi < \theta < 3\pi/2\\ \theta - \pi & 3\pi/2 < \theta < 2\pi\end{cases}\right]
\label{integral_res}
\end{equation}

We determine the parameters ${\cal X}_0$, $\alpha$, and $\theta$ using the maps of Sections~\ref{geometry_Minkowski} and~\ref{bipolar}. For locally Minkowski space, ${\cal X}_0 = 1/A$, and the angles are given by $\alpha =\pi/2$, $ \theta = \pi/2-\delta$. The cusp action is then $S_{\text{cusp}} = -\pi^2M_3/A$. For locally de Sitter space, ${\cal X}_0 = 1/\beta A$, and  $\alpha =\pi/2$, $ \theta = \pi/2-\delta$. The cusp action in this case is then $S_{\text{cusp}} = -\pi^2M_3 /\beta A$.


\begin{thebibliography}{99}

\bibitem{Sauter:1931zz} 
  F.~Sauter,
  ``Uber das Verhalten eines Elektrons im homogenen elektrischen Feld nach der relativistischen Theorie Diracs,''
  Z.\ Phys.\  {\bf 69}, 742 (1931).
  doi:10.1007/BF01339461
  
\bibitem{Gibbons:1986cq} 
  G.~W.~Gibbons,
  ``Quantized flux tubes in Einstein-Maxwell theory and noncompact internal spaces,''
  IN *KARPACZ 1986, PROCEEDINGS, FIELDS AND GEOMETRY 1986*, 597-615 AND PREPRINT - GIBBONS, G.W. (REC.APR.86) 22p

\bibitem{Mellor:1989wc} 
  F.~Mellor and I.~Moss,
  ``Black Holes and Gravitational Instantons,''
  Class.\ Quant.\ Grav.\  {\bf 6}, 1379 (1989).
  doi:10.1088/0264-9381/6/10/008
  
\bibitem{Mann:1995vb} 
  R.~B.~Mann and S.~F.~Ross,
  ``Cosmological production of charged black hole pairs,''
  Phys.\ Rev.\ D {\bf 52}, 2254 (1995)
  doi:10.1103/PhysRevD.52.2254
  [gr-qc/9504015].


  
  
\bibitem{Eardley:1995au} 
  D.~M.~Eardley, G.~T.~Horowitz, D.~A.~Kastor and J.~H.~Traschen,
  ``Breaking cosmic strings without monopoles,''
  Phys.\ Rev.\ Lett.\  {\bf 75}, 3390 (1995)
  doi:10.1103/PhysRevLett.75.3390
  [gr-qc/9506041].
  
\bibitem{Dias:2003st} 
  O.~J.~C.~Dias and J.~P.~S.~Lemos,
  ``Pair creation of de Sitter black holes on a cosmic string background,''
  Phys.\ Rev.\ D {\bf 69}, 084006 (2004)
  doi:10.1103/PhysRevD.69.084006
  [hep-th/0310068].
  
\bibitem{Hawking:1995zn} 
  S.~W.~Hawking and S.~F.~Ross,
  ``Pair production of black holes on cosmic strings,''
  Phys.\ Rev.\ Lett.\  {\bf 75}, 3382 (1995)
  doi:10.1103/PhysRevLett.75.3382
  [gr-qc/9506020].

\bibitem{Emparan:1995je} 
  R.~Emparan,
  ``Pair creation of black holes joined by cosmic strings,''
  Phys.\ Rev.\ Lett.\  {\bf 75}, 3386 (1995)
  doi:10.1103/PhysRevLett.75.3386
  [gr-qc/9506025].

  
\bibitem{Dowker:1993bt} 
  F.~Dowker, J.~P.~Gauntlett, D.~A.~Kastor and J.~H.~Traschen,
  ``Pair creation of dilaton black holes,''
  Phys.\ Rev.\ D {\bf 49}, 2909 (1994)
  doi:10.1103/PhysRevD.49.2909
  [hep-th/9309075].
  
\bibitem{CamposDias:2003tv} 
  O.~J.~Campos Dias,
  ``Black hole solutions and pair creation of black holes in three, four and higher dimensional spacetimes,''
  hep-th/0410294.
  
  
\bibitem{bateman}
BATEMAN MANUSCRIPT PROJECT, BATEMAN, H., \& ERD�LYI, A. (1953). Higher transcendental functions. New York, McGraw-Hill. 

\bibitem{Kinnersley:1970zw} 
  W.~Kinnersley and M.~Walker,
  ``Uniformly accelerating charged mass in general relativity,''
  Phys.\ Rev.\ D {\bf 2}, 1359 (1970).
  doi:10.1103/PhysRevD.2.1359


\bibitem{Anber:2008zz} 
  M.~M.~Anber,
  ``AdS(4) / CFT(3) + Gravity for Accelerating Conical Singularities,''
  JHEP {\bf 0811}, 026 (2008)
  doi:10.1088/1126-6708/2008/11/026
  [arXiv:0809.2789 [hep-th]].
  
\bibitem{Witten:1981mf} 
  E.~Witten,
  ``A Simple Proof of the Positive Energy Theorem,''
  Commun.\ Math.\ Phys.\  {\bf 80}, 381 (1981).
  doi:10.1007/BF01208277
  
\bibitem{Pascu:2012yu} 
  G.~Pascu,
  ``Atlas of Coordinate Charts on de Sitter Spacetime,''
  arXiv:1211.2363 [gr-qc].

\bibitem{Schwinger}
J.~Schwinger,
Phys.\ Rev.\ {\bf{82}}, 664 (1951)
doi:10.1103/PhysRev.82.664

\bibitem{Gibbons:1977mu} 
  G.~W.~Gibbons and S.~W.~Hawking,
  ``Cosmological Event Horizons, Thermodynamics, and Particle Creation,''
  Phys.\ Rev.\ D {\bf 15}, 2738 (1977).
  doi:10.1103/PhysRevD.15.2738






\end{thebibliography}
\end{document}